\documentstyle[prb,aps,epsfig,multicol]{revtex}
\input myfig.sty

\begin{document}

\title{(Meta-)stable reconstructions of the diamond (111)
surface: interplay between diamond- and graphite-like bonding}

\author{A.V. Petukhov$^*$}
\address{RIM and NSR Research Centers, 
Theoretical Physics, University of Nijmegen, 
NL-6525 ED Nijmegen, The Netherlands}
\author{D. Passerone$^{\dag}$, F. Ercolessi, E. Tosatti}
\address{Istituto Nazionale di Fisica della Materia (INFM) and
Scuola Internazionale Superiore di Studi Avanzati (SISSA), 
Via Beirut 2/4, 34014 Trieste, Italy}
\author{A. Fasolino}
\address{RIM Research Center, Theoretical Physics, 
University of Nijmegen, NL-6525 ED Nijmegen, The Netherlands}
\maketitle

\begin{abstract}
Off-lattice Grand Canonical Monte Carlo simulations of the clean 
diamond (111) surface, based on the effective many-body Brenner 
potential, yield the $(2\times1)$ Pandey 
reconstruction in agreement with \emph{ab-initio} 
calculations and predict the existence of new meta-stable states, 
very near in energy, with all surface atoms in three-fold 
graphite-like bonding. We believe that the long-standing debate 
on the structural and electronic properties of this surface 
could be solved by considering this type of carbon-specific 
configurations.
\end{abstract}

\begin{multicols}{2}
The discovery of fullerene has awakened  an increasing interest in carbon 
based nanostructures as well as in 
processes, such as diamond graphitization\cite{graphitization} or 
the graphite-to-diamond transformation\cite{onions}
leading to structures, which promise to have desirable 
properties of both graphite and diamond.
It is important to develop predictive schemes to treat 
diamond, graphite and mixed bonding 
with approaches able to deal with large 
structures, often beyond the possibility of \emph{ab-initio} calculations. 
The effective many-body empirical potentials due to 
Tersoff\cite{Tersoff} for group IV elements 
are very accurate for Si and Ge, 
but less reliable for C. In particular, for C 
the Tersoff potential yields the unreconstructed (111) $(1\times1)$ surface 
as the most stable against the experimental 
evidence\cite{Sowa,Derry,X-Vlieg,MEIS} of a 
$(2\times 1)$ Pandey reconstruction.\cite{Pandey,PandeyC} For the (001) face it 
predicts dimerization with a strong asymmetric displacement of the 
third-layer atoms.\cite{ourPSS} Here, we use the 
potential proposed by Brenner\cite{Brenner} (parametrization I) and 
show that it is reliable also at the surface.\cite{001}

We perform off-lattice Grand Canonical Monte Carlo (GCMC) 
simulations\cite{Celestini} of the (111) surface of diamond. We find the 
unbuckled undimerized $(2\times1)$ Pandey chain reconstruction as the 
minimum energy structure and 
three new meta-stable states, close in energy, with all surface atoms in a 
three-fold graphite-like bonding. Two of them 
are obtained by a strong dimerization of the \emph{lower} 
(4-fold coordinated) chain, inducing a small dimerization of the 
upper ($\pi$-bonded) chain as well. The third meta-stable 
$(\sqrt{3}\times\sqrt{3})R30^{\rm o}$ reconstruction is 
formed by a regular array of vacancies. 

Surprisingly, the reconstruction of  clean diamond(111) is not yet 
established in detail, although there is a consensus that the 
$\pi$-bonded Pandey $(2\times1)$ reconstruction\cite{Pandey,PandeyC} 
is the most stable. One important issue is whether this surface 
is metallic or semiconducting. In most 
calculations\cite{Vanderbilt84,Alfonso,Kern96,Scholze} the band of
surface states is metallic 
whereas experimentally the highest occupied state is at least 0.5 eV below
the Fermi level.\cite{Graupner97,Cui} 
Dimerization along the $\pi$-bonded chain could open the 
surface gap\cite{Tosatti,Bechstedt} but 
only one total-energy calculation obtains slightly dimerized 
chains yielding a 0.3 eV gap\cite{Tosatti} 
in the surface band. Experimentally, 
recent X-ray data\cite{X-Vlieg} and medium-energy 
ion scattering \cite{MEIS} do not show any dimerization but 
favor the $(2\times1)$ 
reconstruction accompanied by a strong tilt of the $\pi$-bonded chains, 
similar to the  $(2\times1)$ reconstruction of Si(111) and Ge(111). 
The tilt is however not confirmed by theoretical 
studies.\cite{Vanderbilt84,Alfonso,Kern96,Scholze,Tosatti} 
Also, relaxations in deeper layers are debated. The bonds between 
first and second bilayers are found to be elongated by an amount which 
varies between 1\%\cite{X-Vlieg,MEIS} and 8\%.\cite{Sowa,Vanderbilt84,Tosatti} 
Bonds between the second and third bilayers are slightly shifted ($\alt1$\%) 
in theoretical studies 
while X-ray data suggest a 5-6\%\ relaxation.\cite{X-Vlieg,MEIS}

Experimentally, uncertainties can be caused by variations of 
surface preparation. A partial 
gra\-phi\-ti\-za\-tion\cite{graphitization,Cui,2x1-or-graphite} 
and other structural phases\cite{Frauenheim,Graupner98} can coexist at the 
real surface. It is noteworthy that most structural 
models were first suggested for Si and 
Ge\cite{Pandey,Chadi,Seiwatz} 
and then extended to diamond.\cite{PandeyC,Vanderbilt84} However, 
the former always favor tetrahedral 
four-fold coordination whereas C 
favors also the graphite-like three-fold bonding, the latter being in 
fact energetically stable in the bulk at normal conditions. One can therefore 
expect diamond to have additional low-energy 
surface structures, such as the meta-stable 
reconstructions presented here. 

Empirical potentials, although less accurate than \emph{ab-initio} 
calculations, allow to explore larger portions of phase space 
and can lead to unexpected structures, which 
can then be tested in more accurate calculations and 
taken into account in the experimental data analysis. 
We exploit the MC simulated annealing scheme to overcome potential 
energy barriers and identify low-energy surface structures. 
Atoms in the bottom layers are kept fixed at their ideal bulk 
positions while the others (usually, four bilayers of 128 atoms each) 
are mobile. We consider 
either a $VNT$, $PNT$ or $P\mu T$ ensembles for different tasks. 
The canonical $VNT$ ensemble is used to minimize 
ordered structures. During the simulated annealing we allow 
for volume fluctuations ($PNT$). 
We consider also a grand canonical $P\mu T$ ensemble\cite{Celestini} 
to access structures 
with different number of atoms than in the bulk terminated structure. 
Each atom creation/destruction is enabled only in the near-surface 
region (2-3 top bilayers) and followed by 1000 MC equilibration moves.
We have optimized the implementation of the potential 
by use of neighbor lists which allow to calculate energy variations 
on a finite portion of the sample. 
The one-dimensional functions defining the potential are stored in 
tables with a fine grid and calculated by linear 
interpolation. The attractive
$V_A$, repulsive $V_R$ and the cut-off $f_c$ terms 
are stored as a function of the square of interatomic distances to 
avoid square root operations. The three-dimensional $F$ function is 
stored on a finer grid and the tricubic interpolation\cite{Brenner} 
is replaced by a linear one, reducing 
the terms to be evaluated from 64 to 8. 
We note that, in a strongly covalent system like diamond, 
creation and destruction are very improbable events, also because 
immediately after creation or destruction the neighboring atoms have 
not yet adjusted to the new local environment; 
after a destruction the system can gain up to 2 eV by 
relaxation. This energy has been added as an 
umbrella\cite{Allen} in the acceptance rule 
for creation/destruction. 

\fone

Throughout, we give energy gains $\Delta E$ per $1\times1$ unit cell, 
relative to the bulk-terminated surface.
We find the  relaxed $(1\times1)$ and Pandey $(2\times1)$ structures
shown in Fig.~\ref{fig1} to have 
$\Delta E=0.244$ and 1.102 eV, respectively.  Apart from 
a 4\%\ elongation of the bond between first and second 
bilayer against 8\%\cite{Sowa,Vanderbilt84,Tosatti} 
for the Pandey structure, our results agree remarkably well with 
\emph{ab-initio} calculations.\cite{Vanderbilt84,Scholze,Tosatti,Stumpf}

To the best of our knowledge, no one has succeeded before in simulating 
a spontaneous transition from the ideal $(1\times1)$ to the 
$(2\times1)$ reconstructed diamond (111) surface. The $(2\times1)$ 
reconstruction of both diamond(111) and Si(111) is associated with a 
large coherent displacement of the first bilayer by more than 0.5 \AA, 
accompanied by rebonding in which 
one atom in the $2\times1$ surface unit cell changes coordination 
from four to three while another does the opposite. 
\emph{Ab-initio} molecular dynamics simulation\cite{Si-spont} 
of the spontaneous $(2\times1)$ reconstruction of Si(111) 
shows that breaking and formation of a new bond occurs at the same 
time. We find that for diamond, instead, the bond breaking leading 
to an increase of three-fold (graphite-like) atoms 
precedes the bond formation.\cite{web} 
In such a situation, competition between the $(2\times1)$ 
reconstruction and (partial) surface graphitization 
becomes very important,\cite{2x1-or-graphite} especially if 
the annealing is performed at high temperatures. Conversely, 
a low annealing temperature makes it very difficult 
to overcome the potential barrier between the two ordered structures. 
In Fig.~\ref{fig2} we show the top view of a diamond (111) sample, obtained
from the ideal relaxed $(1\times1)$ structure after an annealing cycle 
(about $0.5\cdot10^6$ MC steps) at $T=750$ K. 
The efficiency of phase space exploring is improved by increasing 
the step size\cite{Allen} as to reduce the acceptance rate from the 
usual 50\%\ down to 20-25\%. 
In Fig.~\ref{fig2} we emphasize pieces of the lower, four-fold 
coordinated Pandey chains which represent the final stage of the 
$(1\times1) \rightarrow (2\times1)$ transition.  Only one rotational domain 
is present and 
the relative position of the formed chains is somewhat disordered. 

\ftwo

\fthree
\tabl

At higher temperatures, apart from a tendency towards 
graphite-like structures, we observe also other 
ordered phases, similar to the Pandey reconstruction but accompanied 
by a strong dimerization of the \emph{lower} atomic chain.\cite{web} 
This dimerization can be performed in two ways, 
leading to the $(2\times1)$ and $(4\times1)$ structures shown in 
Fig.~\ref{fig3}. The atom coordinates are given\cite{web} in 
Table~\ref{table} along with those for the Pandey 
reconstruction. The energy 
gain $\Delta E$ is $0.883$ and $1.023$ eV, 
for the metastable $(2\times1)$ and $(4\times1)$ respectively. 
In the dimerized chains the short/long distances between atoms 
13 and 14 are 1.459/2.215 and 1.444/2.455 \AA\ for the $(2\times1)$ 
and $(4\times1)$ instead of 1.562 \AA\ in the Pandey reconstruction. 
This dimerization induces dimerization of the $\pi$-bonded chain 
as well, albeit small ($<1$\%), which might be of importance for surface 
electronic properties. The $(4\times1)$ dimerized structure is only 
160 meV ($\approx 2000$ K) 
per broken bond higher than the Pandey structure, i.e. 
these phases can coexist at the surface at high temperatures. 
Indeed, by heating the ordered Pandey 
structure to $2350$ K we observe a partial transformation to the 
dimerized state 
as shown in Fig.~\ref{fig4}. Note also a precursor of surface 
graphitization. 

%\fthree
%\tabl

Lastly, in GCMC runs we find the $(\sqrt{3}\times\sqrt{3})R30^{\rm o}$
reconstruction formed by an ordered array of vacancies as shown on 
the right hand side in Fig.~\ref{fig3}. The bond lengths between 
the atoms in the first bilayer is reduced to 1.390 \AA, 
slightly less than the equilibrium bond length in graphite 
(1.42 \AA). The bonds between the first and second layer are elongated 
by $\sim 1$\%\ while the other bond lengths are close to the bulk 
value. 
Taking the bulk binding energy 7.346 eV as the chemical potential, 
the energy gain is estimated to be $\Delta E=0.6145$ eV. Once formed, 
this structure is found to be very stable and remains unaltered after long  
GCMC annealing cycles at $T=2350$ K.\cite{web} 
Similar structures have been discussed for  
Si.\cite{sq3-exp,sq3-th} For diamond(111) 
a $(2\times2)$ vacancy structure was shown to be energetically unfavorable 
compared to the relaxed $(1\times1)$ structure.\cite{Bechstedt2} 
However, contrary to the $(2\times2)$ vacancy 
structure,\cite{Bechstedt2} our 
$(\sqrt{3}\times\sqrt{3})$ structure has only three-fold 
coordinated surface atoms and might thus be more favorable. 

\ffour

In conclusion, we have performed off-lattice GCMC simulations of 
the clean diamond (111) surface structure based on the Brenner potential. 
A spontaneous transition from the 
ideal $(1\times1)$ to the stable Pandey $(2\times1)$ reconstruction 
is obtained. We also find metastable reconstructions very 
close in energy with strong dimerization of the lower atomic chain, 
which are shown to coexist with the Pandey chain reconstruction at temperatures 
$\sim 2000$ K. Besides, we find a deep local minimum for the vacancy 
stabilized $(\sqrt{3}\times\sqrt{3})$ structure. These meta-stable 
structures have a larger number of three-fold coordinated atoms 
at the surface. The absence of consensus on 
the structural details and electronic structure of the clean (111) 
surface might be related to these surface structures, which 
are peculiar of carbon and have never been considered so far. 

%Acknowledgments: 
A.V.P. and A.F. would like to thank E. Vlieg, F. van Bouwelen, 
J.J. ter Meulen, W. van Enckevort, J. Schermer and B.I. Dunlap for 
useful discussions. D.P., F.E. and E.T. acknowledge support by MURST.

\end{multicols}
\end{document}